\documentstyle {article}
\newcommand{\bqq}{\begin{equation}}
\newcommand{\eqq}{\end{equation}}
\newcommand{\rrr}{roughening}

\begin{document}

\begin{center} {\LARGE {\bf Kinetic roughening in growth models with
diffusion in higher dimensions }}\\

\vspace{10pt}

{\large P. \v{S}milauer $^{a,}$\footnote{\noindent On leave from Institute
of Physics, Academy of Sciences, Cukrovarnick\'{a} 10, 162 00 Praha 6,
Czech Republic}
and M. Kotrla $^{b}$}\\
\vspace{5pt}
$^a$  {\it Interdisciplinary Research Centre for Semiconductor Materials,
Imperial College, London SW7 2BZ, United Kingdom}\\
$^b$ {\it Institute of Physics, Academy of Sciences, Na Slovance 2,\\ 180
40 Praha 8, Czech Republic}\\

\vspace{15pt}

{\large {\bf Abstract}}\\
\end{center}

\noindent We present results of numerical simulations of kinetic \rrr\ for
a growth model with surface diffusion (the Wolf-Villain model) in 3+1 and
4+1~dimensions using lattices of a linear size up to $L\!=\!64$ in 3+1~D
and $L\!=\!32$ in 4+1~D. The effective exponents calculated both from the
surface width and from the height--height correlation function are much
larger than those expected based on results in lower dimensions, due to a
growth instability which leads to the evolution of large mounded
structures on the surface. An increase of the range for incorporation of a
freshly deposited particle leads to a decrease of the roughness but does
not suppress the instability.

\vspace*{30pt}

Kinetic roughening in growth models has been extensively studied in recent
years \cite{rev}. Two approaches are mainly used: a macroscopic, continuum
theory based on stochastic differential equations, and numerical
simulations of discrete models. The width (roughness) of a growing surface,
$w(t,L)\!=\!\langle\sqrt{\overline{h^2}\!-\!\overline{h}^{2}}\rangle$ (the
variance of the surface height profile $h({\bf x},t)$), obeys the
dynamical scaling law $w(t,L)\propto L^{\zeta} f(t/L^{z})$ as a function
of time $t$ and the system size $L$, where $f(y)\rightarrow$ const as
$y\!\rightarrow\!\infty$ and $f(y)\propto y^{\beta}, \beta\!=\!\zeta/z$ as
$y\!\rightarrow\!0$. The asymptotic behavior of the surface width is
characterized by the exponents $\zeta$ and $z$ (or $\zeta$ and $\beta$)
which depend on the universality class of the model used and on the
spatial dimension $d\!=\!d'\!+\!1$, $d'$ being the surface dimension.

Recently, a new class of models with diffusion (or, more precisely,
post--deposition relaxation) has been investigated \cite{villain,wv,lai}. A
physical mechanism behind the surface smoothing in these growth models is
surface diffusion controled by the nearest--neighbor bonding energy and it
was supposed that these simplified models should mimic the situation in
growth by molecular beam epitaxy. The effective scaling exponents obtained
from simulations of such models are often compared to the exponents
obtained using approximate analytical methods \cite{villain,lai,tang} from
differential equations of the form
\bqq
\frac{\partial h({\bf x},t)}{\partial t} =  -\nabla \cdot {\bf j}({\bf
x},t) + \eta ({\bf x},t) \label{eq:1}
\eqq
where the current ${\bf j}({\bf x},t)$ is a function of the derivatives of
$h({\bf x},t)$. (Note, however, that the connection between microscopic
computer models and continuum equations is far from being
straightforward.) Wolf and Villain (WV) \cite{wv} formulated a discrete
growth model in which a particle arriving on the surface relaxes in a
local neighborhood to increase the number of bonds to nearest neighbors.
If there is no possibility to increase the number of nearest neighbors,
the particle stays at the initial position. First numerical simulations of
the WV model in 1+1~D \cite{wv} yielded exponents in surprising agreement
with those of the linear differential equation (${\bf j} \propto
\nabla\nabla^2 h$). However, new extensive simulations \cite{ks3} revealed
a crossover to intermediate behavior with the exponents
corresponding to a  nonlinear equation, ${\bf j} \propto \nabla(\nabla h)^2$,
and indicated a further crossover to Edwards--Wilkinson behavior (${\bf j}
\propto-\nabla h$), the second crossover being  in agreement with recent
calculations by Krug et al. \cite{krug} using the inclination--dependent
diffusion current.\footnote{Recent results show that the asymptotic
behavior of the WV model is of the Edwards--Wilkinson type. However, the
correct description of the intermediate behavior is not known at the
moment, in particular since the values of the exponents $\zeta^{\rm c}$
calculated at this regime from the correlation function (see below)
correspond to an equation with a nonlinear term $\nabla(\nabla h)^3$ in
1+1~D, but to an equation with a nonlinear term $\nabla^2(\nabla h)^2$ in
2+1~D and cannot be explained simultaneously by any of the models
considered so far \cite{ks3}.} Simulations in 2+1~D \cite{kls,ks3} gave
exponents in agreement with those obtained from the nonlinear equation
($\zeta\!=\!(5-d)/3$, $\beta\!=\!(5-d)/(7+d)$). Note that simulations of a
full--diffusion (FD) model, in which all surface atoms can move not only
immediately after deposition, but during the whole simulation with the
rates determined by nearest--neighbor bonding \cite{wvz}, yielded
exponents of the nonlinear equation in 1+1, 2+1, {\it and\/}
3+1~D.\footnote{The WV model and the FD model are of course different. The
Laplacian term should not be present in the FD model equation where the
hopping probability depends only on the initial coordination. However, we
have found that in 1+1 dimensions there is very close correspondence in
behavior of both models for quite a long time interval with the radius of
incorporation
$R_i$ (described below) corresponding to the temperature in the FD model.}

In order to understand the situation better and to check the universality
class of the model, we have performed extensive simulations of the WV
model in higher unphysical dimensions. Below, we present our results in
3+1 and 4+1~D. The model under study is a straightforward generalization
of the model investigated in the lower dimensions \cite{wv,ks3,kls}. A
particle is deposited at a randomly selected site on the surface and then
relaxes to a position with a highest {\it coordination\/} (the number of
nearest neighbors) $v$, according to the following rules. The particle
relaxes to a site with the highest coordination $v_{\rm max}$ of all
the sites examined if this is also higher than the coordination $v_{\rm
init}$ of the initial site of incidence, $v_{\rm max}\!>\!v_{\rm init}$.
If there are several sites with $v\!=\!v_{\rm max}\!>\!v_{\rm init}$, a
random choice is made. If there is no site with a coordination higher than
$v_{\rm init}$, {\it i.e.\/} $v_{\rm max}\!=\!v_{\rm init}$, the
particle stays at the initial position. The solid--on--solid condition is
obeyed in the model, {\it i.e.\/}, no overhangs or bulk vacancies are
allowed. At first we have restricted the possible sites for the relaxation
to nearest neighbors only.

The simulations in higher dimensions are much more demanding on computer
time not only because of more degrees of freedom but also due to very slow
dynamics of the model (the dynamical exponent predicted for, e.g., the
nonlinear model, $z\!=\!(7+d)/3$, is large and increases with the spatial
dimension). Therefore, one has to deposit a large number of layers to
reach the stationary state characterized by the saturated surface width.
The maximal linear system sizes for which we were able to measure the
saturated surface roughness were $L\!=\!45$ (we can only estimate it for
$L\!=\!64$) in 3+1~D and $L\!=\!16$ in 4+1~D. Exponent $\beta_{\rm eff}$
was calculated using data for the largest systems, $L\!=\!64$ (3+1~D) and
$L\!=\!24$ (4+1~D). Due to the enormous computer time required we averaged
over several independent runs only.

In contrast to the situation in 1+1~D  and 2+1~D \cite{ks3,kls}, we have
found in the higher dimensions unexpectedly rapid roughening with the
effective exponent $\beta_{\rm eff}$ showing an initial increase with time
(Fig. 1). In 3+1~D, $\beta_{\rm eff}$ increases from 0.12 to
$\approx\!0.3$ (Fig. 2) which is much larger than the value predicted by
any of the above--mentioned continuum equations. We have analyzed our data
using a linear fit to 7 successive points on the double logarithmic scale
(Fig.~2). Maximal values of the effective exponent $\beta_{\rm eff}(L)$
for different system sizes $L\!=\!32-64$ are still slightly increasing
which indicates that the size $L\!=\!64$ is still not large enough to
measure the correct value of the effective exponent. In Fig.~1, the
dependence of the saturated surface width on the system size is shown as
well. The exponent $\zeta_{\rm eff}$ increases with the system size from
$\zeta_{\rm eff}\!\approx\!0.47$ for $L\!=\!4-8$ to $\zeta_{\rm
eff}\!\approx\!1.25$
for $L\!=\!32-45$; this is again much larger than the values predicted by
any of the continuum equations. It seems that after the initial increase
for $L\!=\!4-16$, $\zeta_{\rm eff}$ remains approximately constant. A
linear fit for $L\!=\!16-64$ yields $\zeta_{\rm eff}\!\approx\!1.2$. The
corresponding value of the dynamical exponent $z$ is about $4$. One can
also estimate $z$ from the system--size dependence of the time after which
the surface width saturates. In this way we obtain $z\!\approx\!3.6$. We
observed similar behavior even in 4+1~D which is predicted by some
of the macroscopic theories to be the upper critical dimension. In this
case $\beta_{\rm eff}(L)\!\approx\!0.25$ for $L\!=\!32$ and $\zeta_{\rm
eff}\!\approx\!0.94$ for $L\!=\!8-16$.

Another way of studying the kinetic \rrr\ is the calculation of the
height--height correlation function $G({\bf r},t)\!=\!\langle [h({\bf
x}+{\bf r},t)-h({\bf x},t)]^2\rangle$. Recently, it has been
found \cite{schroeder,ks3} that the WV model in 1+1 and 2+1 D exhibits
anomalous scaling behavior and the exponents calculated from the surface
width and from the correlation function are different. We observed this
anomaly related to the increase of the average step height also in 3+1~D
and 4+1~D. The exponents $\zeta^{\rm c}_{\rm eff}$ obtained from the
correlation function at later times (when $\beta_{\rm eff}$ is large) are:
$\zeta^{\rm c}_{\rm eff}\!=\!0.81\!\pm\!0.03$ in 3+1~D and $\zeta^{\rm
c}_{\rm eff}\!=\!0.7\!\pm\!0.1$ in 4+1~D. The sudden increase in the value
of $\beta_{\rm eff}$ described above is reflected in a faster power--law
increase of the average step height $G(1,t)\!\propto\!t^\lambda$ with
$\lambda\!\approx\!0.08$. However, unlike in 1+1 and 2+1~D
\cite{ks3,schroeder}, obtained values of
$\zeta$, $\zeta^{\rm c}$ and $\lambda$ are inconsistent with the
relation suggested in Ref. \cite{schroeder}. In other words, this
anomalous behavior caused by the increase of the average step height does
not explain the large values of the effective exponents which are
instead caused by a mounding instability described below which leads to
breakdown of conventional scaling and an evolution of a macroscopically
corrugated surface.

Because of the unexpected results of our simulations, we have carefully
checked possible sources of errors. The same behavior in 3+1~D was
observed using different random number generators, and even using a
different code \cite{mw}. We also tried to obtain more information on the
evolution of the system and studied the statistics of jumps of freshly
deposited particles in different directions (jumps up and down are the
most important for the developing roughness). We have found that the
probability of jumps in different lateral directions is well isotropic. The
number of jumps up monotonously increased up to the region of the saturated
roughness where it saturated. The only quantity which seems to be
related to the increase in $\beta_{\rm eff}$ is the number of jumps down,
which at first rapidly increases and then, for several tens of monolayers,
has a maximum and after that slowly decreases.

An important information has been provided by the study of the surface
morphology using 2+1--dimensional sections of the 3+1--dimensional
$L\!=\!50$ lattice. We observed a development of mound structures. Since
the time $2^{10}$ small mounds are seen in different places which later
increase in size and merge so that finally there is only one large mound
in the whole sample (a typical example is shown in Fig.~3). The
development of such mounds suggests the existence of an instability which
we studied by calculations of a slope--dependent currents for tilted
surfaces ($h\!\rightarrow\!h+{\bf x \cdot m}$) following the paper by Krug
et al. \cite{krug}. Similarly to 1+1 and 2+1~D (see \cite{krug}), we found
a current in the {\it downhill\/} direction, of the order of
$10^{-4}$ for the slopes $m\!=\!2$ and $m\!=\!1$. However, for small slopes
($m\!=\!0.5$ and $m\!=\!0.25$) we observed in contrast to the lower
dimensions an {\it uphill\/} current, which causes the instability
observed (detailed results will be published elsewhere). The situation is
similar to growth with the Schwoebel barrier to hopping down step edges
where for small slopes a destabilizing uphill current which increases with
the slope is observed whereas it monotonously decreases (but still remains
uphill) for large slopes \cite{johnson}. However, in our case the current
changes its sign and becomes downhill for large slopes. In fact, we
observed that the current is very close to zero for simulations on a
surface with the average slope of $m\!=\!0.75$. Visual inspection of
mounds similar to the one shown in Fig.~3 reveals that their angle of
inclination is very close to the value 0.75, in agreement with the results
of the work of Krug {\it et al.\/} \cite{krug} according to which a slope
with the zero diffusion current is selected. We do not understand, however,
what drives the uphill current and why such behavior is not observed in
1+1 and 2+1 dimensions and these questions remain to be answered.

To study the problem further, we also performed simulations of two
modifications of the WV model. At first we studied the modified WV model
in which the deposition site is chosen not only from nearest--neighbor
sites but from all the sites in a region of a linear dimension $R_i$
around the site of incidence.

We used the algorithm employed earlier in the simulation of the FD model for
the relaxation of incoming particles \cite{clarke}. In this case, a site
on the surface is selected randomly, and a site with the maximum
coordination within a cube with a side of $2R_i\!+\!1$ lattice constants
centered upon the original site is chosen as the site of final deposition
(notice that $R_i\!=\!1$ is different from the original WV model since also
next--nearest neighbors are considered). We have observed that the
roughness saturates much earlier for a larger $R_i$ and the value of the
saturated roughness progressively decreases with increasing $R_i$. This
makes it difficult to measure the exponent $\beta_{\rm eff}$ for large
$R_i$'s; larger and larger sizes $L$ are needed (Fig.~4). However, within
our error bars, we have obtained the same large $\beta_{\rm
eff}\!\approx\!0.3$ for $R_i\!=\!1,2,3,5$. One could see that the region
with an initial small value of $\beta_{\rm eff}$ disappears with
increasing $R_i$. The exponent $\zeta_{\rm eff}$ ($L\!=\!40-60$) decreases
with increasing $R_i$ from $\zeta_{\rm eff}\!\approx\!1.18$ for $R_i\!=\!1$
to $\zeta_{\rm eff}\!\approx\!0.69$ for $R_i\!=\!5$. Similarly, the
exponent $\zeta_{\rm eff}^{\rm c}$ obtained from the height--height
correlation function decreases with increasing $R_i$ from $\zeta_{\rm
eff}^{\rm c}\!\approx\!0.82$ for $R_i\!=\!1$ to $\zeta_{\rm eff}^{\rm
c}\!\approx\!0.5$ for $R_i\!=\!3$. The decrease of the surface width is
on the microscopic level reflected by the increase in the number of jumps
down (and the disappearance of the small initial $\beta_{\rm eff}$ by the
disappearance of the maximum in the number of jumps down). We also studied
another modification of the WV model in which jumps up are forbidden
(model II of Ref. \cite{kls}) and observed a similar behavior with an
increase to a large $\beta_{\rm eff} \approx 0.25$ after $\approx 10^2$
layers were deposited. In both considered modifications, the roughness is
lower than in the original WV model, but the mounding instability is still
present.

In conclusion, we have carried out extensive simulations of a growth model
proposed by Wolf and Villain with local post--deposition relaxation in 3+1
and 4+1~D. We have found effective exponents much larger than those
expected based on results for 1+1~D and 2+1~D. Anomalous scaling
due to the power--law increase of the average step height observed in 1+1~D
and 2+1~D (manifested in different values of the roughness exponent
$\zeta_{\rm eff}$ obtained from the surface width evolution and
$\zeta^{\rm c}_{\rm eff}$ from the height--height correlation function) is
also present in the higher dimensions. However, a dominant role is played
by a growth instability resulting in a macroscopically modulated surface
profile with large mounds developing on the surface. Calculations of
diffusion currents for tilted surfaces show a destabilizing uphill current
at small slopes microscopic origin of which remains unclear. This current
changes its sign (becomes downhill) as the slope increases and our results
suggest that the angle of inclination of the mounds observed on the surface
corresponds to the slope for which the current is zero. An increase in the
radius of incorporation of a freshly arrived particle leads to a decrease
of the roughness but does not suppress the instability. More work is
needed to determine why the mounding instability is observed here but not
in the lower dimensions.

Work performed at Imperial College is supported by Imperial College and the
Research Development Corporation of Japan. P.~\v{S}. would like to thank
Dr. M.R.~Wilby for valuable discussions and suggestions on the model and
simulation technique, and for communicating his results prior to
publication. M.~K. wishes to thank Prof.~A.C.~Levi for valuable
discussions.

\newpage

\noindent {\bf\large Figure captions}\\
Fig.1 - Surface width $w$ {\it vs.} time $t$ for the WV model (relaxation
to nearest neighbors only) in 3+1~D. Solid squares $L\!=\!64$, stars
$L\!=\!45$, solid triangles $L\!=\!32$, open squares $L\!=\!16$, open
circles $L\!=\!8$, open triangles $L\!=\!4$. Inset: saturated surface width
$w_{\infty}$ {\it vs.\/} system size for the same model.\\
Fig.2 - Evolution of the effective exponents $\beta_{\rm eff}$ for the WV
model in 3+1 D for different system sizes. Solid squares $L\!=\!64$, stars
$L\!=\!45$, solid triangles $L\!=\!32$.\\
Fig.3 - Example of the surface morphology (a 2+1~D section of a 3+1~D
lattice) obtained for a lattice of the linear dimension $L\!=\!50$ after
$2^{16}$ monolayers were deposited.\\
Fig.4 - Comparison of the time evolution of the surface width in the
modified WV model with the different incorporation radii
$R_i\!=\!1,2,3,5$ in 3+1~D.

\end{document}